\documentclass[runningheads,12pt]{colloq}
\usepackage[latin1]{inputenc}
\usepackage[french]{babel}

\usepackage{color}
\definecolor{red}{rgb}{0.75,0,0}
\definecolor{blue}{rgb}{0,0,0.75}
\definecolor{green}{rgb}{0,0.75,0}

\usepackage{graphicx,epsf,subfigure}  % standard LaTeX graphics tool
\def\approxgt{\,\raise2pt \hbox{$>$}\kern-8pt\lower2.pt\hbox{$\sim$}\,}
\def\approxlt{\,\raise2pt \hbox{$<$}\kern-8pt\lower2.pt\hbox{$\sim$}\,}

\def\ih{{\^\i }}

\begin{document}
 
\title*{Du chaos dans la musique des \'etoiles}
 
%\titretab{début du titre
%\protect\newline fin du titre}
% permet un saut de ligne dans la table des matières
\titretab{Du chaos dans la musique des sph\`eres}
 
\titrecourt{Du chaos dans la musique des sph\`eres}
% permet une abréviation du titre pour les hauts de page
 
\author{\large J. Robert Buchler\inst{1}
\and \large Zoltan Koll\'ath\inst{2}}

\auteurcourt{J. Robert Buchler \& Zoltan Koll\'ath}

% si il y a plus de deux auteurs, mettez le nom du premier auteur suivi de 
% {\it et al}.
 
\adresse{Physics Department, University of Florida
         Gainesville FL32611
\and Konkoly Observatory, Budapest, Hungary
}

\email{buchler@phys.ufl.edu, kollath@buda.konkoly.hu,
http://www.phys.ufl.edu/$\sim$buchler.}

\maketitle              

\begin{abstract}
La plupart des \'etoiles variables qui pulsent avec grande amplitude, telles
les c\'eph\'eides, ont un comportement tr\`es r\'egulier et p\'eriodique.  Mais
juste \`a c\^ot\'e d'elles dans le diagramme \hyphenation{Hertz-sprung-Russ-ell}
Hertzsprung-Russell, se trouve un groupe d'\'etoiles variables dot\'ees de
courbes de lumi\`ere tr\`es irr\'eguli\`eres.  A l'aide d'une technique de
reconstruction de flot appliqu\'ee aux donn\'ees d'observation astronomiques de
plusieurs de ces \'etoiles, on montre que la dynamique sous-jacente est
chaotique et de faible dimension, ce qui peut surprendre \`a cause de la
violence de ces pulsations.  En plus \`a l'aide d'une lin\'earisation du flot
on d\'eduit que le m\'ecanisme physique de la pulsation consiste en
l'interaction entre deux modes vibratoires (vraisemblablement radiaux), l'un
lin\'eairement instable, de fr\'equence $f_0$, et l'autre stable, mais de
fr\'equence $\sim 2f_0$ (soit un sc\'enario \`a la Shilnikoff
g\'en\'eralis\'e).\\
\ \\
{\normalfont \bf Colloque sur le "Chaos temporel et chaos spatio-temporel",
Septembre 2001, Le Havre, FRANCE}
\end{abstract}

\section{Introduction}

L'\'energie \'emise par les \'etoiles est tr\`es souvent non constante, ceci
pouvant \^etre pour des raisons extrins\`eques ou
intrins\`eques.  L'obscurcissement p\'eriodique d'une \'etoile par un compagnon
binaire, ou l'\'emission intermittante de gaz qui se condensent et
obscurcissent l'\'etoile comptent parmi le premier type de variabilit\'e
stellaire.  Dans le second type de pulsation on distingue en plus, d'un
c\^ot\'e entre excitation stochastique de modes de pulsation comme dans le
soleil, pulsations qui pour la plupart du temps concernent un grand nombre de
modes non-radiaux et de tr\`es faible amplitude, et d'un autre c\^ot\'e les
pulsations auto-excit\'ees, le plus souvent radiales et de grande amplitude
o\`u seulement un ou deux modes prennent part.  Pour ce dernier type de
pulsation c'est l'interaction entre la pulsation et le flux de chaleur qui est
responsable pour l'instabilit\'e vibratoire \cite{cox}.  Dans cette revue c'est
ce dernier type de variabilit\'e qui nous int\'eresse, et il s'av\`ere tr\`es
utile d'\'etudier ces \'etoiles pulsantes dans le cadre des syst\`emes
dynamiques \cite{buchler93}.
         
Les \'etoiles pulsantes les mieux connues sont les c\'eph\'eides classiques et
les \'etoiles dites du type RR~Lyrae.  Ces \'etoiles variables sont
p\'eriodiques avec des cycles limites d\^us \`a l'excitation d'un seul mode
lin\'eairement instable, qui peut \^etre soit le mode fondamental, soit le
premier ou deuxi\`eme overtone\footnote{Nous utilisons le mot anglais pour
\'eviter une confusion possible, car les modes ne forment pas une s\'equence
harmonique, multiples d'une fr\'equence fundamentale.}.  Les c\'eph\'eides \`a
bosse ('bump cepheids') sont caract\'eris\'ees par l'excitation synchronis\'ee
d'un deuxi\`eme mode lin\'eairement stable, mais r\'esonnant dans un rapport de
fr\'equence 2:1.  Dans le cas des c\'eph\'eides dites ``c\'eph\'eides \` a deux
modes'' les pulsations sont bi-p\'eriodiques dans le sens qu'il y a deux modes
incommensurables excit\'es \`a des amplitudes constantes.  A cause de la
fameuse relation p\'eriode-luminosit\'e les c\'eph\'eides p\'eriodiques ont
jou\'e et continuent de jouer un r\^ole essentiel dans la d\'etermination de
l'\'echelle cosmologique.

Cette revue ne s'int\'eresse cependant pas aux c\'eph\'eides classiques, mais
plut\^ot aux \'etoiles variables dot\'ees de courbes de lumi\`ere
irr\'eguli\`eres qui, dans le diagramme Hertzsprung-Russell (diagramme
luminosit\'e -- temp\'erature effective), occupent une bande assez large et
courb\'ee, situ\'ee en dessous des c\'eph\'eides classiques \cite{cox}.  Pour
des raisons historiques et astronomiques les \'etoiles dans cette bande ont des
appellations diff\'erentes, malgr\'e que leurs propri\'et\'es changent
graduellement le long de cette bande.  Ainsi les \'etoiles qui se trouvent aux
plus faibles luminosit\'es sont appel\'ees \'etoiles du type W~Virginis.
Au-dessus se trouvent les \'etoiles du type RV~Tauri, suivies des \'etoiles
dites semi-r\'eguli\`eres et finalement les \'etoiles du type Mira.  Dans ce
qui suit nous appellerons toutes ces \'etoiles semi-r\'eguli\`eres, {\it largo
sensu}.  Les \'etoiles W~Vir au d\'ebut de cette bande ont des courbes de
lumi\`ere p\'eriodiques ($P\approxlt 20d$), mais au-del\`a elles pr\'esentent
des alternances dans les cycles qui ont toutes les apparences d'un doublement de
p\'eriode \cite{arp55}. En fait les mod\'elisations num\'eriques exhibent une
cascade de doublement de p\'eriode \cite{b87}, \cite{pdchaos88}, \cite{aik90}
ainsi que des bifurcations tangentielles \cite{bgk87}, \cite{aik87}.  Les
observations montrent aussi que l'irr\'egularit\'e des courbes de lumi\`ere de
ces \'etoiles cro\ih t avec leur luminosit\'e \cite{pol96}.

\begin{figure}[htbp]
  \begin{center}
    \epsfysize=20cm
    \leavevmode
    \epsfbox{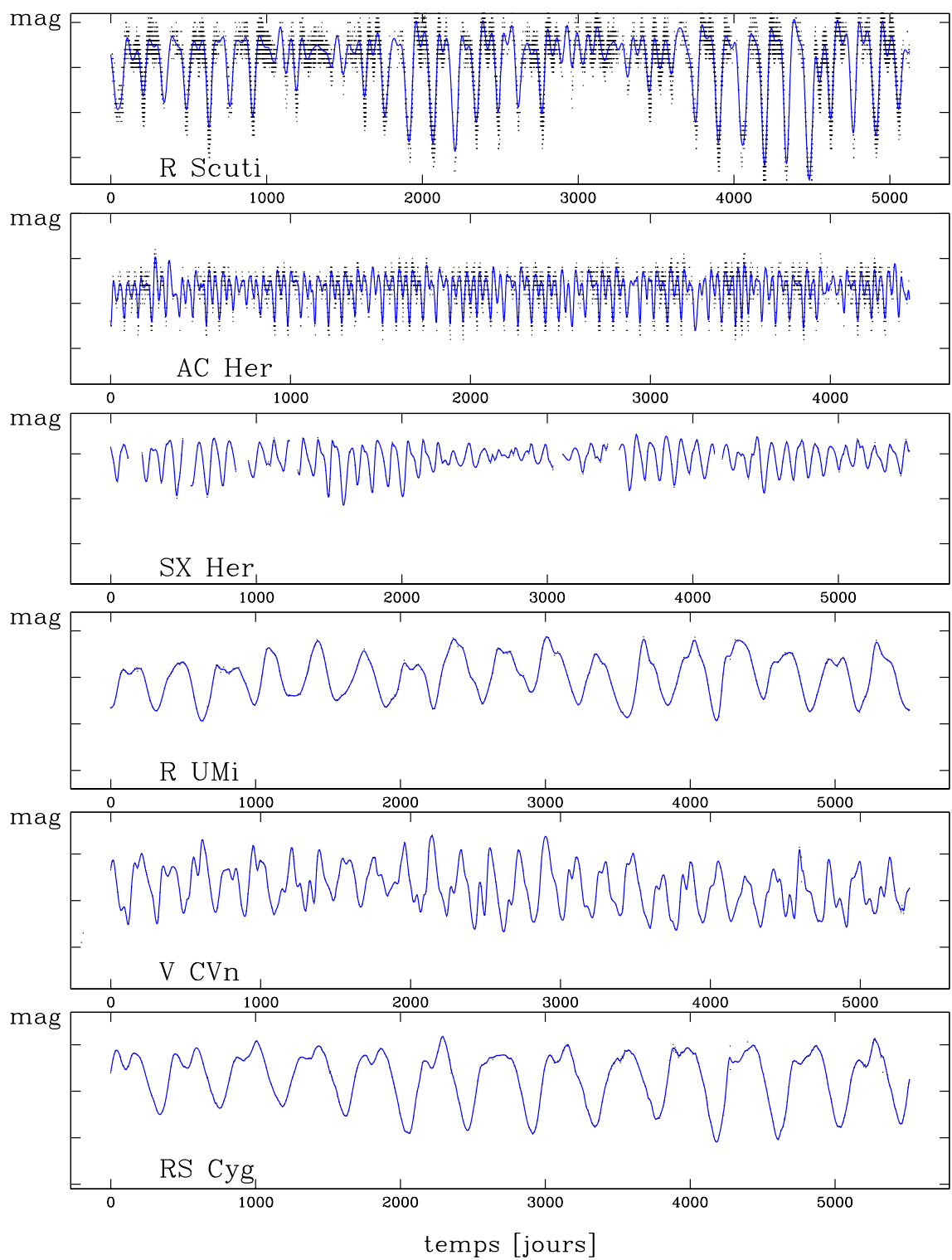}
    \caption{Courbes de lumi\`ere observationnelles liss\'ees superpos\'ees sur
  les donn\'ees d'observation}
    \label{fig_LC}
  \end{center}
\end{figure}

Il y a quelques ann\'ees nous avons eu acc\`es \`a une base de donn\'ees de
deux \'etoiles variables irr\'eguli\`eres, s'appelant R~Scuti et AC~Herculis,
donn\'ees prises par des astronomes amateurs et compil\'ees par l'AAVSO
(American Association of Variable Star Observers.  Nous renvoyons aux
r\'ef\'erences \cite{rsctprl} \cite{rsct96} \cite{acher98} pour notre analyse
de ces donn\'ees.  Le d\'esavantage est que ces donn\'ees ne sont pas tr\`es
pr\'ecises, mais par contre le recouvrement temporel est excellent, et il y a
tr\`es peu de trous.  Plus r\'ecemment, gr\^ace aux efforts observationnels de
R. Cadmus \cite{potsdam} nous avons pu faire une analyse pr\'eliminaire des
courbes de lumi\`ere de quatre \'etoiles suppl\'ementaires, R~Ursae Minoris, SX
Herculis, RS Cygni et V Canum Venaticorum.  Dans notre analyse nous avons
liss\'e les donn\'ees observationelles astronomiques par spline cubique avec
$\sigma=0.02$ \cite{reinsch}) et nous avons utilis\'e un \'echantillonnage
\'equidistant d'un jour.  Les courbes de lumi\`ere des six \'etoiles sont
exhib\'ees dans la figure~\ref{fig_LC}.

\section{La reconstruction de flot}

Notre analyse consiste essentiellement en une reconstruction de flot avec une
forme polynomiale \`a plusiers variables.  Cette m\'ethode s'est av\'er\'ee
aussi tr\`es utile dans des applications autres qu'astronomiques (voir par
exemple la revue de \cite{letellier-floride}).  En astronomie elle a aussi
\'et\'e utilis\'ee pour l'analyse des cycles solaires \cite{serre}.

La m\'ethode qui remonte \`a Farmer et Packard, Crutchfield et Farmer et \`a
Takens (voir \cite{weigend}) est toute simple en principe.  A l'aide de la
s\'erie temporelle scalaire $\{x_i\}$ des donn\'ees observationnelles rendues
\'equidistantantes par lissage et interpolation on construit un vecteur
d'\'etat $X\in \bbR^{d_e}$, tel que

\begin{equation}
{\bf X}^n = (x_i, \thinspace x_{i-\tau}, \thinspace  x_{i-2 \tau}, \thinspace  
\ldots  \thinspace  x_{i-(d_e-1) \tau} )
\end{equation}

On fait l'hypoth\`ese que le signal \`a \'etudier est g\'en\'er\'e par
une dynamique de faible dimension, et par cons\'equent on postule
l'existence d'une application  ${\cal M}$ dans un espace de
plongement de dimension $d_e$ telle que

\begin{equation}
{\bf X}^{n+1} = {\cal M} ({\bf X}^n).  
\end{equation}
L'application ${\cal M}$ est donn\'ee comme somme de tous les monomes
construites avec les composantes de ${\bf X}$, donc de la forme

$${\cal B}_k = ({\bf X}^n)_{j_1} ({\bf X}^n)_{j_2} ({\bf X}^n)_{j_3} \ldots$$
avec $j_1 + j_2 + j_3 \ldots = k$, $k\le p$ form\'es \`a partir des composantes
du vecteur ${\bf X}$ jusqu'\`a un ordre $p$.  Les coefficients des monomes sont
ensuite calcul\'es par la m\'ethode des moindres carr\'es (effectu\'ee de
mani\`ere num\'eriquement stable \`a l'aide d'une d\'ecomposition en valeurs
singuli\`eres (SVD)).  Souvent, en pratique, il suffit d'aller jusqu'\`a
l'ordre $p=4$.

\begin{figure}[htbp]
  \begin{center}
    \epsfysize=9cm
    \leavevmode
  \epsfbox{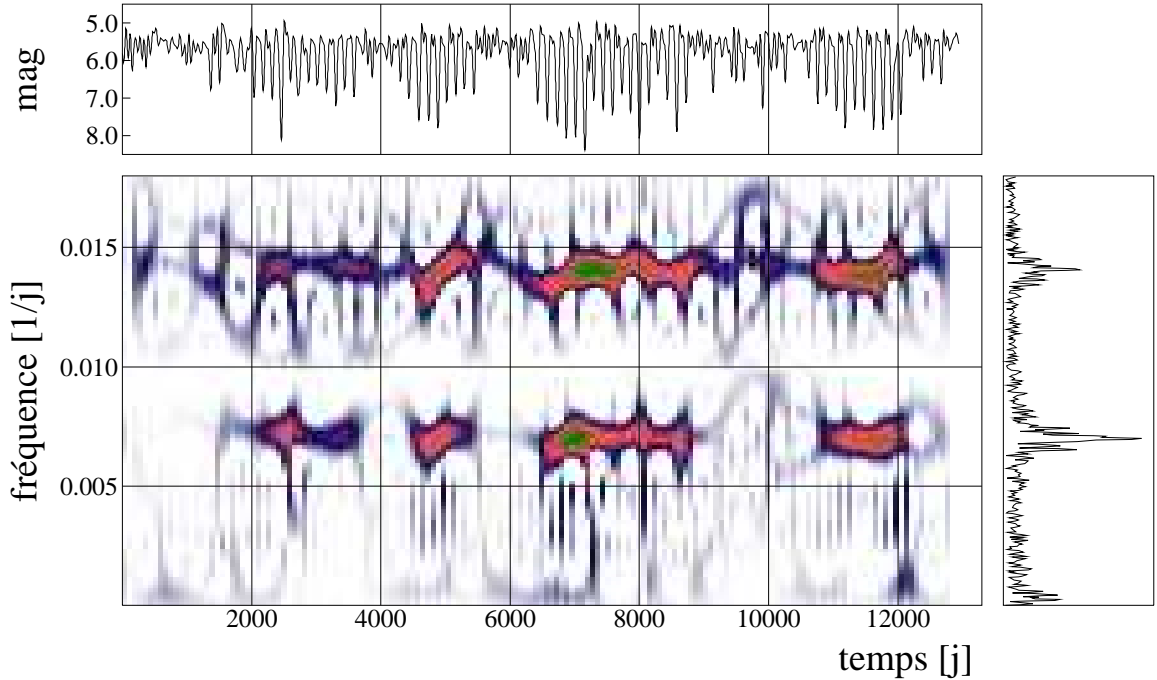}
%  \epsfbox{rsc_con.ps}
%%  \epsfbox{rsc_gab.ps}
%%  \epsfbox{rsc_cwd.ps}
 \vskip 1pt
   \caption{En haut: Courbe de lumi\`ere observationnelle liss\'ee de R~Sct;
     \`a droite: spectre de Fourier (amplitude); 
     centre:  diagramme temps-fr\'equence.   }
    \label{rsct}
  \end{center}
\end{figure}

Quand la s\'erie observationnelle est donn\'ee sur des intervalles de
temps suffisamment courts pour qu'on puisse approximer la d\'eriv\'ee par
des diff\'erences (p. ex. \`a la Adams-Moulton) on peut alors calculer
un flot ${\bf F}$

\begin{equation}
{d{\bf X}\over dt} = {\bf F}({\bf X}).
\end{equation}
D'apr\`es notre exp\'erience cette derni\`ere reconstruction est en
g\'en\'eral moins robuste.

Notre reconstruction d\'epend de trois param\`etres qui sont \`a priori
arbitraires: la dimension de l'espace de plongement $d_e$, l'ordre
maximal des monomes $p$, ainsi que le param\`etre de d\'elai $\tau$.  Ce
dernier doit \^etre assez long pour que la reconstruction ne soit pas
noy\'ee dans le bruit, mais assez courte pour que le flot ou
l'application ne deviennent pas trop nonlin\'eaires \cite{serre96}.
Nous disons que notre reconstruction est robuste quand nous obtenons
des r\'esultats semblables sur un intervalle contigu de plusieurs
valeurs de $\tau$.

Pour plus de d\'etails nous r\'ef\'erons le lecteur int\'eress\'e \`a
deux articles de revue \cite{varenna} et \cite{takeuti}, et pour les
d\'etails techniques de nos reconstructions \`a \cite{serre96},
\cite{rsctprl}, \cite{rsct96}, \cite{acher98}.

Ici nous nous contenterons de montrer les r\'esultats.

\begin{figure}[htbp]
  \begin{center}
    \epsfysize=8cm
    \leavevmode
    \epsfbox{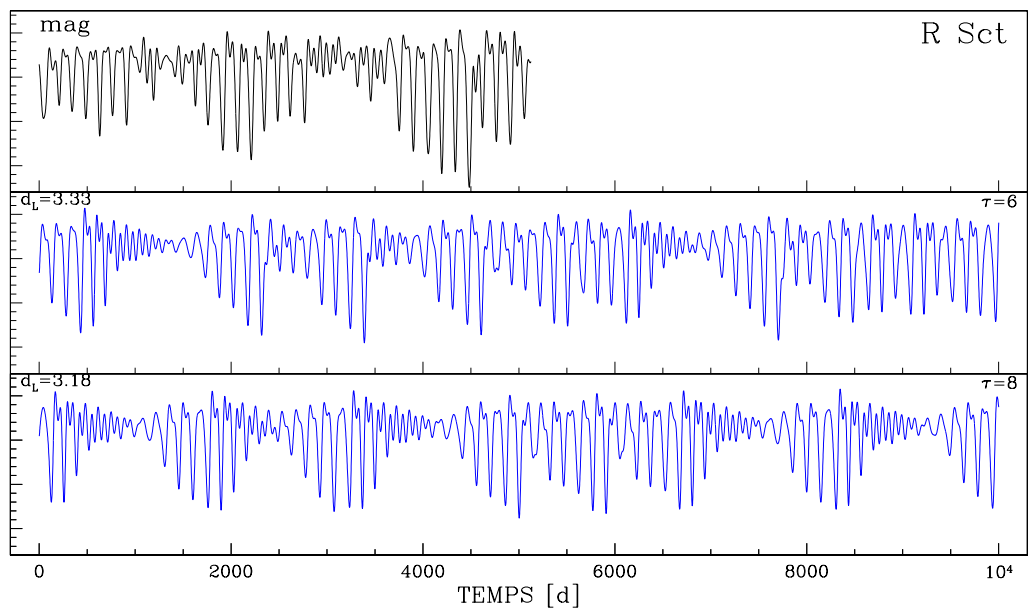}
    \caption{De haut en bas: Courbes de lumi\`ere observationnelle (magnitudes)
             liss\'ee de R~Sct; courbe de lumi\`ere synth\'etique typique.  A
             gauche est indiqu\'ee la dimension fractale $d_L$ et \`a
             droite le d\'elai $\tau$ utilis\'e.}
    \label{rsct_syn}
  \end{center}
\end{figure}

\subsection{L'\'etoile R Scuti}

Dans la figure \ref{rsct} nous montrons en plus de la courbe de lumi\`ere (en
magnitudes) liss\'ee, le spectre de Fourier amplitude et un diagramme
temps-fr\'equence, fait avec un noyau \`a la Cohen \cite{cohen}.  (Pour le
non-astronome, pour des raisons historiques la magnitude $m$ est d\'efinie
comme le logarithme de $L$, viz. $m = -2.5 \thinspace {\rm Log} L$).  On
constate que le spectre pr\'esente un pic large vers 35 jours, ainsi qu'un pic
tr\`es large et approximativement harmonique vers 70 jours.  L'analyse
temps-fr\'equence montre que les frequences sont variables, et dans le spectre
de Fourier les sous-pics individuels ne sont pas stables -- ils varient d'une
section de donn\'ees sur l'autre et ceci de mani\`ere non r\'eguli\`ere
(voir\cite{kollath90}).  La courbe de lumi\`ere n'est donc pas
multi-p\'eriodique, mais varie de mani\`ere nonstationnaire.  On note le
changement de puissance aussi dans le diagramme temps-fr\'equence.
 
On a pu montrer que la courbe de lumi\`ere de l'\'etoile R~Scuti, du type
\'etoile RV~Tau a \'et\'e g\'en\'er\'ee par une dynamique \`a quatre
dimensions.  Ces conclusions sont bas\'ees sur les faits suivants:\hfill\break
(a) La dimension de plongement minimale est $d_e=4$: ceci sort de la
comparaison des courbes de lumi\`ere synth\'etiques avec les donn\'ees
observationnelles; ces signaux synth\'etiques sont obtenus par l'it\'eration
de l'application ${\cal M}$ qu'on a constuite \`a partir des donn\'ees par la
m\'ethode des moindres carr\'es.  A $d_e=3$ les courbes synth\'etiques n'ont
aucune ressemblance avec les donn\'ees; deuxi\`emement, l'erreur du fit

\begin{equation} E = \sum_n ||{\bf X}^{n+1} - {\bf F}({\bf X}^n)||
\end{equation} atteint un plateau pour $d_e=4$; troisi\`emement, la m\'ethode
des "voisins les plus proches" \cite{brown} indique \'egalement $d_e=4$ comme
dimension minimale.  L'\'evidence la plus forte vient de la comparaison des
spectres de Fourier des signaux synth\'etiques avec celles des donn\'ees
observationlles, ainsi que leur comparaison au niveau des projections \`a la
Broomhead-King sur les vecteurs propres de la matrice de corr\'elation.

Les reconstructions sont tr\`es robustes, donnant essentiellement les m\^emes
r\'esultats pour des valeurs de $\tau$ allant de 4 \`a 8 .  Cela ne veut
cependant pas dire que pour toute valeur de d\'epart dans l'it\'eration on
puisse produire une courbe de lumi\`ere synth\'etique, ni qu'on puisse it\'erer
ind\'efiniment; pour certaines valeurs l'it\'eration diverge tout de suite,
pour d'autres elle peut le faire apr\`es un million d'it\'erations ou plus.  Il
est clair qu'avec une courbe de lumi\`ere observationnelle de dur\'ee aussi
courte que celle qu'on analyse ici on n'explore pas tout l'espace de phase de
la dynamique, et il existe des endroits faiblement couverts o\`u le flot
reconstruit diverge.

Les courbes de lumi\`ere synth\'etiques reproduisent aussi bien les alternances
de minima faibles et profonds que les modulations de dur\'ee plus longue.  Pour
cette \'etoile les reconstructions sont assez robustes pour produire des
it\'erations suffisamment longues ($>100,000$ points) pour que la valeur des
exposants de Lyapunoff ($\{ \lambda_k \}$ et de la dimension $d_L$ soient
stables num\'eriquement.

Nous montrons une courbe de lumi\`ere typique dans la figure
\ref{rsct_syn}.  Il est peut-\^etre int\'eressant de mentionner que ce
genre de pulsation a \'et\'e observ\'e depuis 150 ans pour R~Sct
(\cite{kollath90}).  
%\vfill\break

(b) La question s'impose s'il est \`a priori \'evident qu'il existe une
dynamique sous-jacente qui puisse \^etre captur\'ee par un flot de faible
dimension.  Pour un flot (autonome) il doit toujours exister un exposant de
Lyapunoff $\lambda_k$ \'egal \`a z\'ero.  Pour une application discr\`ete on s'attend donc
\`a un exposant proche de z\'ero quand l'intervalle de temps entre les $x_i$
est suffisamment petit (pour que l'application soit proche d'un flot continu).
Or nous trouvons toujours qu'entre le premier exposant positif (indiquant la
pr\'esence de chaos) et le troisi\`eme exposant n\'egatif, il existe bien un
exposant proche de z\'ero.  Ceci confirme que nous capturons la dynamique pour
cette \'etoile.

(c) La dimension fractale $d_L$ est d\'efinie \cite{ott} comme

\begin{equation}
d_L =  N +  \sum_{k=1}^N {\lambda_k\over |\lambda_{N+1}}| 
\end{equation}
o\`u $N$ est le plus grand indice pour lequel la somme des exposants de
Lyapunoff $\lambda_k$ (par ordre d\'ecroissant) est positive.  Nous trouvons
que les $d_L$ des signaux synth\'etiques tombent dans l'intervalle $3.1 - 3.2$
pour R~Sct.  Ce qui est \'evidemment rassurant c'est que la dimension fractale
n'augmente pas avec la dimension de plongement.  Il est \'evident que la
dimension physique, c'est-\`a-dire le nombre de coordonn\'ees physiques
(variables de l'espace de phase) n\'ecessaires pour la description de la
dynamique, et donc la dimension de l'espace de phase physique $d$ est
sandwich\'e entre la dimension fractale et la dimension de plongement minimale

\begin{equation}
 3.2 \approx d_L < d \le d_e^{min} = 4
\end{equation}

(d) Le r\'esultat qui est peut-\^etre le plus int\'eressant du point de vue
physique vient de la lin\'earisation des applications autour de leur points
fixes: on trouve deux racines lin\'eaires du type spirale, $\pm i \omega +
\xi$, avec les propri\'et\'es suivantes: $\omega_2 \approx 2 \omega_1$,
$\xi_1<0$.  Ce sc\'enario de Shilnikoff g\'en\'eralis\'e \cite{tresser}
pr\'edit la pr\'esence de chaos dans un large domaine de param\`etres.

A partir de ce dernier r\'esultat nous pouvons donner une interpr\'etation
physique \`a la dynamique.  Les c\'eph\'eides classiques dans le domaine de
p\'eriode de 7 \`a 15 jours ('bump cepheids') ex\'ecutent des oscillations
p\'eriodiques dans leur mode fondamental dans lequel le second 'overtone' est
entra\ih n\'e par une r\'esonance du type 2:1 entre les fr\'equences
d'oscillation \cite{buchler93}.  On voit que l'\'etoile R~Scuti en fait autant:
les amplitudes complexes de ces deux modes peuvent \^etre consid\'er\'ees comme
des coordonn\'ees g\'en\'eralis\'ees de l'espace de phase physique de dimension
4.  La dynamique appara\ih t comme l'interaction nonlin\'eaire de deux modes
vibratoires, le premier lin\'eairement instable et le second, lin\'eairement
stable et en r\'esonance 2:1 approximative avec le premier.  Il s'ensuit des
alternances chaotiques de croissance et de d\'ecroissance.

Le lecteur peut se demander pourquoi dans les cas des c\'eph\'eides classiques le
m\^eme ph\'enom\`ene de r\'esonance donne lieu \`a des oscillations
p\'eriodiques, alors que dans le cas de R~Sct on aboutit avec des oscillations
chaotiques.  La raison est que malgr\'e que les c\'eph\'eides classiques aient la
m\^eme luminosit\'e que leurs cousines, les \'etoiles type RV~Tau, la masse de
ces derni\`eres est 10 fois plus faible, ce qui entra\ih ne que le couplage
entre l'oscillation et le flux de chaleur dans l'\'etoile est beaucoup plus
grand.  Si nous appelons $\sigma = \pm i \Omega + \kappa$ \thinspace les
valeurs propres des oscillations lin\'eaires, nous trouvons que les taux de
croissance $|\kappa/\Omega |$, de l'ordre du pourcent pour les c\'eph\'eides
atteignent des valeurs d'ordre unit\'e pour les RV~Tau.  Ceci est \'evidemment
une condition n\'ecessaire pour l'existence de modulations d'amplitude (chaos)
sur une \'echelle de temps dynamique ($\Omega^{-1}$).

\begin{figure}[htbp]
  \begin{center}
    \epsfysize=9cm
    \leavevmode
    \epsfbox{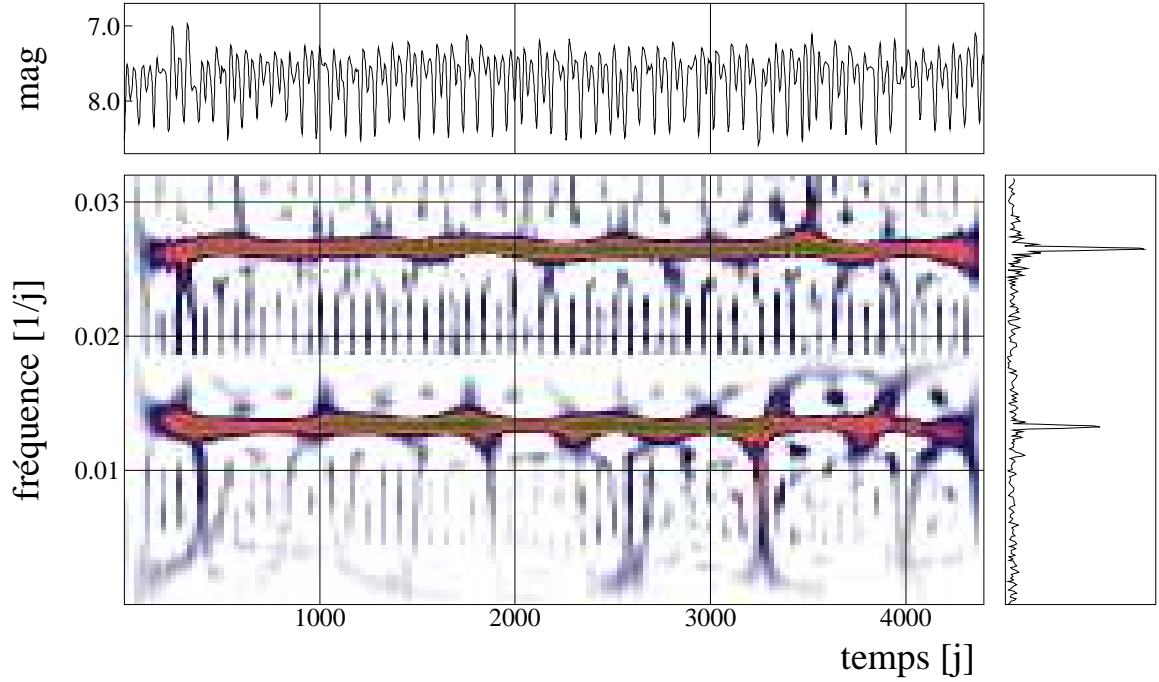}
%    \epsfbox{acher_cwd.ps}
 \vskip 1pt
    \caption{En haut: Courbe de lumi\`ere observationnelle liss\'ee de AC~Her;
     \`a droite: spectre de Fourier (amplitude); 
     centre:  diagramme temps-fr\'equence.   }
    \label{acher}
  \end{center}
\end{figure}

\begin{figure}[htbp]
  \begin{center}
    \epsfysize=8cm
    \leavevmode
    \epsfbox{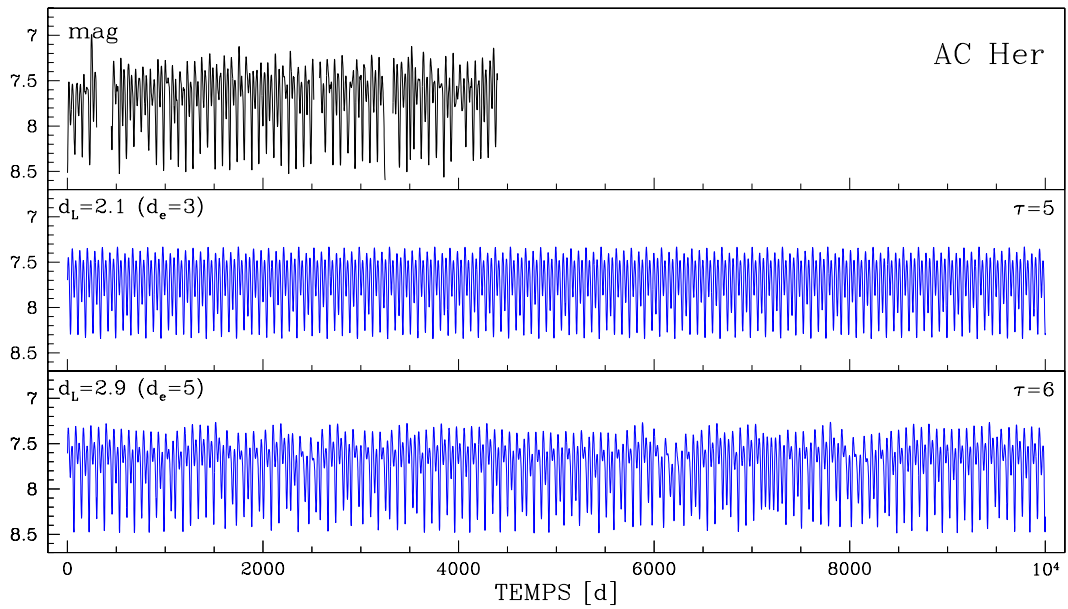}
    \caption{De haut en bas: Courbes de lumi\`ere observationnelle (magnitudes)
             liss\'ee de AC~Her; courbe de lumi\`ere synth\'etique typique.  A
             gauche est indiqu\'ee la dimension fractale $d_L$ et \`a
             droite le d\'elai $\tau$ utilis\'e.}
    \label{acher_syn}
  \end{center}
\end{figure}

\subsection{L'\'etoile AC~Herculis}

La courbe de lumi\`ere liss\'ee, le spectre de Fourier de l'amplitude et le
diagramme temps-fr\'equence sont montr\'es dans la figure \ref{acher}.
L'analyse de cette \'etoile, \'egalement du type RV~Tau, est un bien moins
d\'efinitive. En particulier, les courbes synth\'etiques \`a 3D sont plus
r\'eguli\`eres (Fig.\ref{acher_syn}) que les observations, il n'y en a pas \`a
4D.  La raison des difficult\'es r\'eside probablement dans la petitesse du
rapport signal-bruit.  La dimension de plongement minimale $d_e^{min}$ est entre
3 et 5.

La lin\'earisation des applications autour du point fixe n'est pas robuste, ce
qui n'\'etonne pas \'etant donn\'e que le signal n'a pas de phases de faible
amplitude et par cons\'equent n'explore pas le voisinage de l'origine.  M\^eme
si $d_e^{min}=4$ sugg\`ere de nouveau la pr\'esence de deux modes
oscillatoires, on ne peut donc pas confirmer la pr\'esence d'une r\'esonance
comme dans R~Sct.

\begin{figure}[htbp]
  \begin{center}
    \epsfysize=9cm
    \leavevmode
    \epsfbox{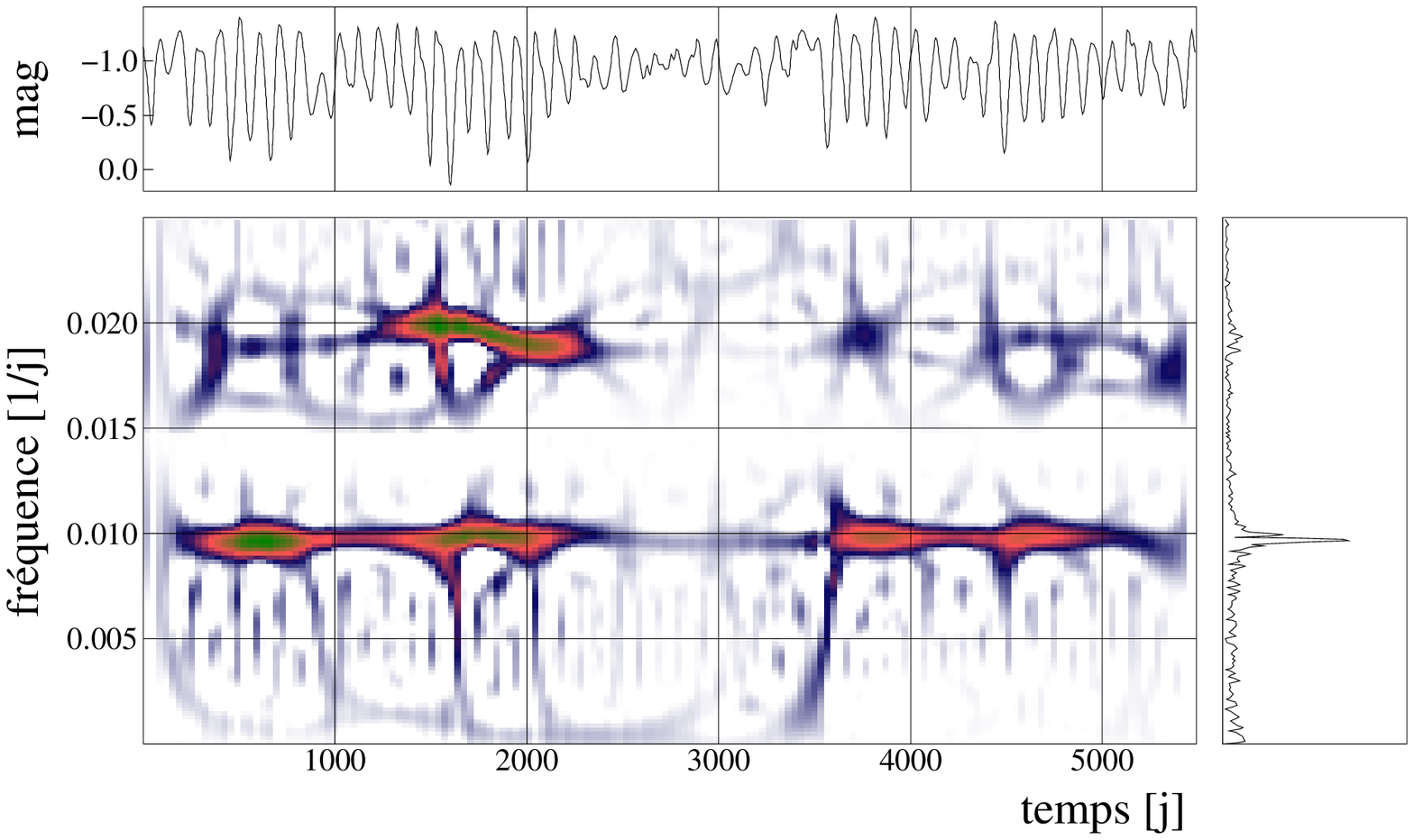}
%    \epsfbox{sxher3.ps}
 \vskip 1pt
    \caption{En haut: Courbe de lumi\`ere observationnelle liss\'ee de SX~Her;
     \`a droite: spectre de Fourier (amplitude); 
     centre:  diagramme temps-fr\'equence.   }
    \label{sxher}
  \end{center}
\end{figure}

\begin{figure}[htbp]
  \begin{center}
    \epsfysize=8cm
    \leavevmode
    \epsfbox{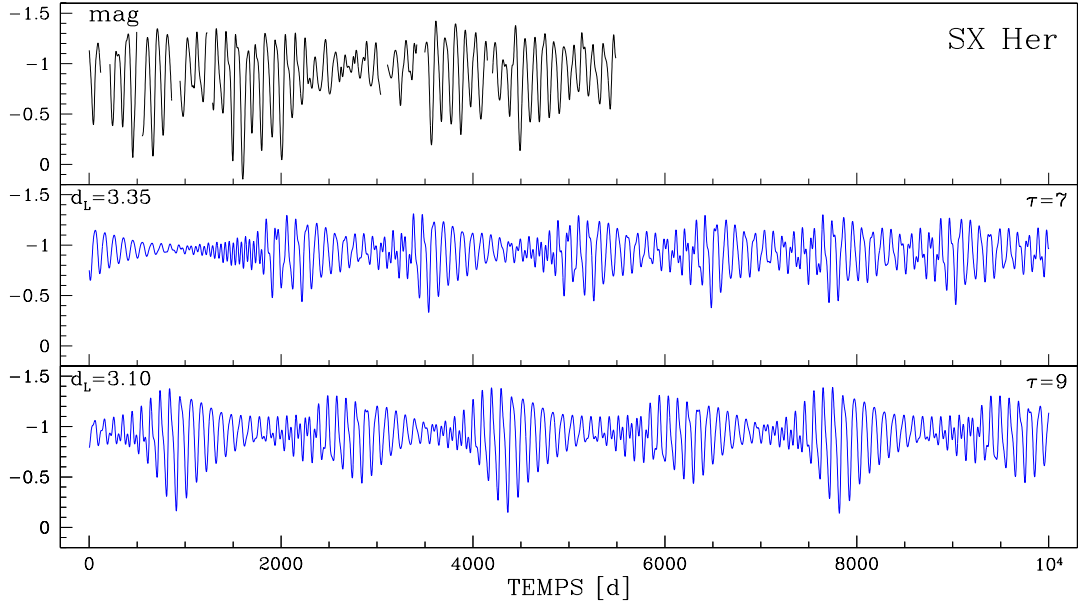}
    \caption{De haut en bas: Courbes de lumi\`ere observationnelle (magnitudes)
             liss\'ee de SX~Her; courbe de lumi\`ere synth\'etique typique.  A
             gauche est indiqu\'ee la dimension fractale $d_L$ et \`a
             droite le d\'elai $\tau$ utilis\'e.}
    \label{sxher_syn}
  \end{center}
\end{figure}

\subsection{L'\'etoile SX Herculis}

La courbe de lumi\`ere liss\'ee, le spectre de Fourier de l'amplitude et le
diagramme temps-fr\'equence sont montr\'es dans la figure \ref{sxher}.  Dans le
spectre de Fourier la puissance est concentr\'ee dans un pic fundamental large
avec un peu de puissance dans la premi\`ere harmonique.

Les reconstructions avec les courbes de lumi\`ere (en magnitude) ne sont pas
possibles \`a 3 dimensions ($d_e=3$) .  A quatre dimensions par contre, elles
capturent les traits principaux des observations comme le montre la figure
\ref{sxher_syn}.  Les reconstructions ne sont pas aussi robustes que celles de
R~Sct, et en fait il est difficile de produire des signaux synth\'etiques de
plus de 10,000 points avant que l'it\'eration diverge; en plus cela n'a \'et\'e
possible que pour deux valeurs de $\tau=7$ et 9.  Les dimensions fractales
(incertaines \`a cause de la courte dur\'ee des signaux synth\'etiques) vont de
3.1 \`a 3.7 dans 4D.  A 5D les reconstructions sont difficiles et peu robustes.
Cependant nous notons que les quelques signaux synth\'etiques que nous ayons pu
produire \`a 5D avaient tous une dimension fractale plus petite que 3.  Nous en
concluons, de mani\`ere tentative, que pour cette \'etoile aussi la dimension
de l'espace de phase physique est egalement $d=4$.

\begin{figure}[htbp]
  \begin{center}
    \epsfysize=8cm
    \leavevmode
    \epsfbox{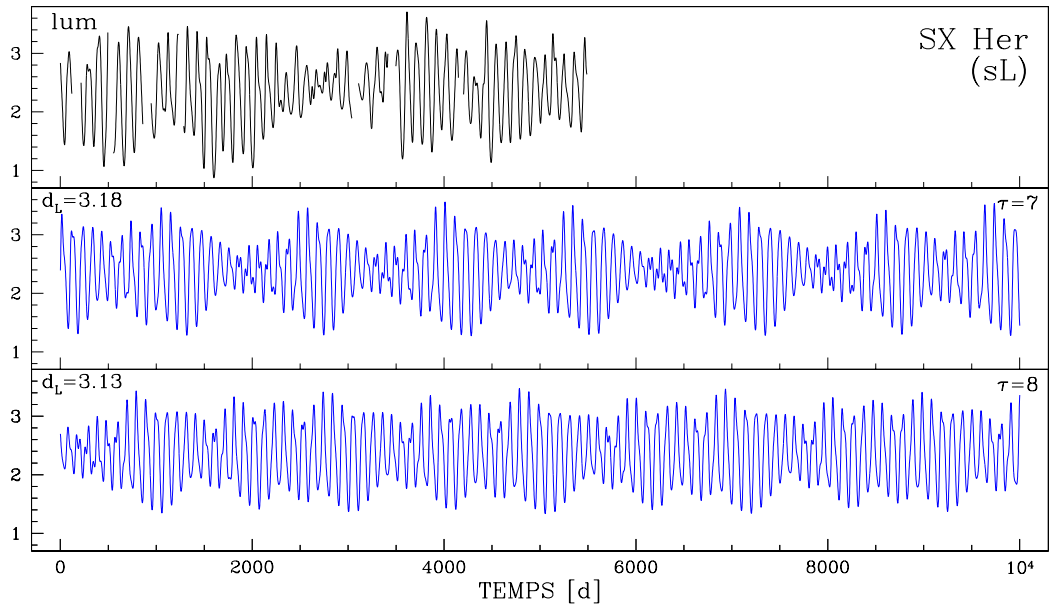}
    \caption{De haut en bas: Courbes de lumi\`ere (en luminosit\'e)
             observationnelle liss\'ee de SX~Her; courbe de lumi\`ere synth\'etique
             typique.  A gauche est indiqu\'ee la dimension fractale
             $d_L$ et \`a droite le d\'elai $\tau$ utilis\'e.}
    \label{sxher_syn_sL}
  \end{center}
\end{figure}

Quant \`a la lin\'earisation de l'application autour du point fixe (qui est
situ\'e au centre du mouvement) nous trouvons aussi 2 racines spirales \`a
l'instar de R~Sct: Le rapport des deux fr\'equences est de 2.1 -- 2.2, et le
taux de croissance (la partie r\'eelle de la racine) est plus grand pour le
mode instable que pour le mode stable, si bien qu'on est de nouveau en
pr\'esence d'un sc\'enario \`a la Shilnikoff g\'en\'eralis\'e

Comme les reconstructions avec les magnitudes ne sont pas aussi satisfaisantes
que nous aurions pu l'esp\'erer, nous avons \'egalement fait des
reconstructions avec les courbes de lumi\`ere (en luminosit\'e) qui sont
montr\'ees dans la figure \ref{sxher_syn}.  On constate que les signaux
synth\'etiques (\`a 4D) sont meilleurs qu'avec les magnitudes, m\^eme si les
reconstructions n'ont \'et\'e possibles qu'avec 2 valeurs de $\tau=7$ et 8.  Il
est int\'eressant de noter cependant que les propri\'et\'es physiques sont les
m\^emes: les dimensions fractales qu'on trouve vont de 3.2 \`a 3.4, c-\`a-d
qu'elles sont plus grandes que 3 et moindres que 4.  Les points fixes sont du
type spirale avec des rapports de fr\'equences proches de et l\'eg\`erement
plus grands que 2.  Nous notons cependant qu'il n'a pas \'et\'e possible de
faire des reconstructions stables \`a 5D.

Nous en concluons qu'il est probable que la dynamique de cette \'etoile est
tr\`es semblable \`a celle de R~Sct, et r\'esulte donc de l'interaction de deux
modes de pulsation, l'un de fr\'equence $f_o$, lin\'eairement instable, qui
entra\ih ne par effet de r\'esonance un second mode stable de fr\'equence
$\approx 2 f_o$.

\begin{figure}[htbp]
  \begin{center}
    \epsfysize=9cm
    \leavevmode
    \epsfbox{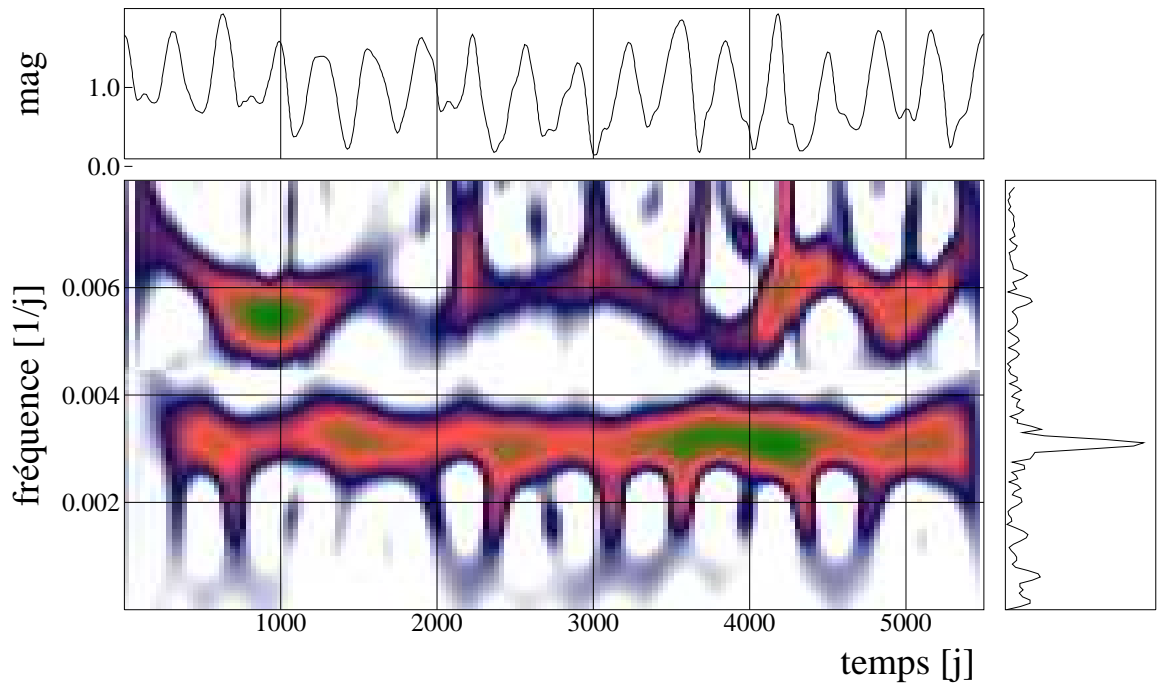}
%    \epsfbox{rumi2.ps}
 \vskip 1pt
    \caption{En haut: Courbe de lumi\`ere observationnelle liss\'ee de R~UMi;
     \`a droite: spectre de Fourier (amplitude); 
     centre:  diagramme temps-fr\'equence.   }
    \label{rumi}
  \end{center}
\end{figure}

\begin{figure}[htbp]
  \begin{center}
    \epsfysize=8cm
    \leavevmode
    \epsfbox{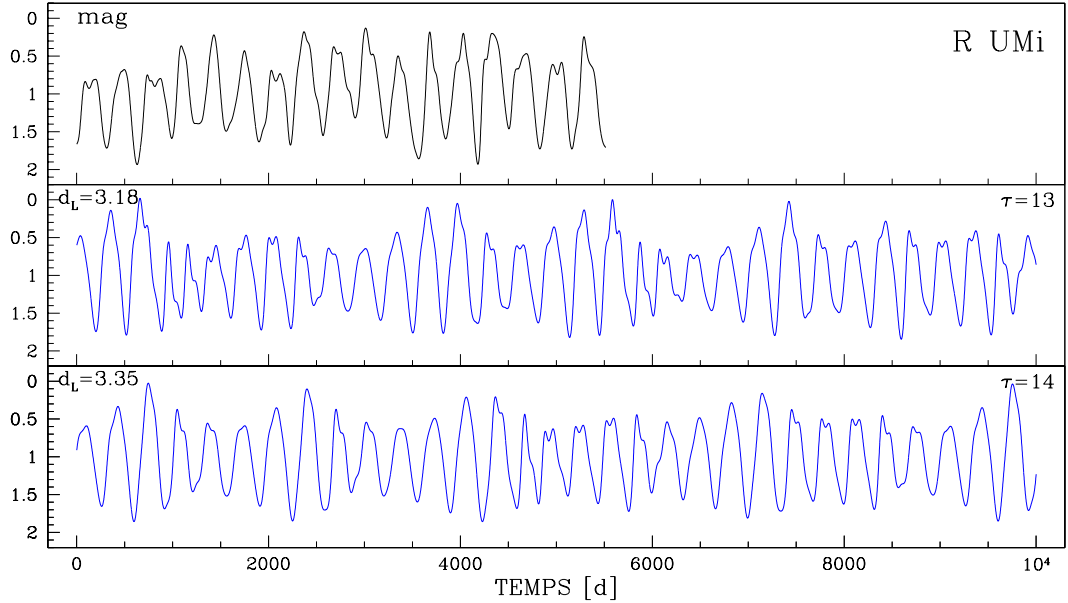}
    \caption{De haut en bas: Courbes de lumi\`ere observationnelle (magnitudes)
             liss\'ee de R~UMi; courbe de lumi\`ere synth\'etique typique.  A
             gauche est indiqu\'ee la dimension fractale $d_L$ et \`a
             droite le d\'elai $\tau$ utilis\'e.}
    \label{rumi_syn}
  \end{center}
\end{figure}

\subsection{L'\'etoile R~Ursae Minoris}

La courbe de lumi\`ere liss\'ee, le spectre de Fourier de l'amplitude et le
diagramme temps-fr\'equence sont montr\'es dans la figure \ref{rumi}.  La
courbe de lumi\`ere de cette \'etoile est vraiment de dur\'ee tr\`es courte en
ce qui concerne le nombre de cycles, et on pourrait croire qu'une
reconstruction soit condamn\'ee d'office.  Cependant, il a \'et\'e possible de
faire des reconstructions assez robustes, pour des valeurs de $\tau$ allant de
3 \`a 10 jours (quoique pour quelques valeurs interm\'ediaires ce ne f\^ut pas
possible).  A cause de la courte dur\'ee du signal observationnel les signaux
synth\'etiques sont en g\'en\'eral de courte dur\'ee (10,000 points) et les
exposants de Lyapunoff ne sont pas d\'etermin\'es avec beaucoup de pr\'ecision.
Nous trouvons que les dimensions fractales des signaux synth\'etiques
acceptables tombent autour 3.25 $\pm$ 0.15 pour $\tau\ge 8$.  Il est tr\`es
rassurant que cette dimension est stable dans le sens que les reconstructions
\`a 5D donnent des $d_L$ du m\^eme ordre, en particulier plus petites que 4.
(Nous notons cependant qu'il semble y avoir une autre dynamique plus
r\'eguli\`ere avec $d_L = 3.1 - 3.4$ pour $\tau = 3 - 6$.)  A 3D les
reconstructions ne sont pas satisfaisantes, ce qui sugg\`ere que la dimension
$d_e^{min} = 4$.

\begin{figure}[htbp]
  \begin{center}
    \epsfysize=9cm
    \leavevmode
    \epsfbox{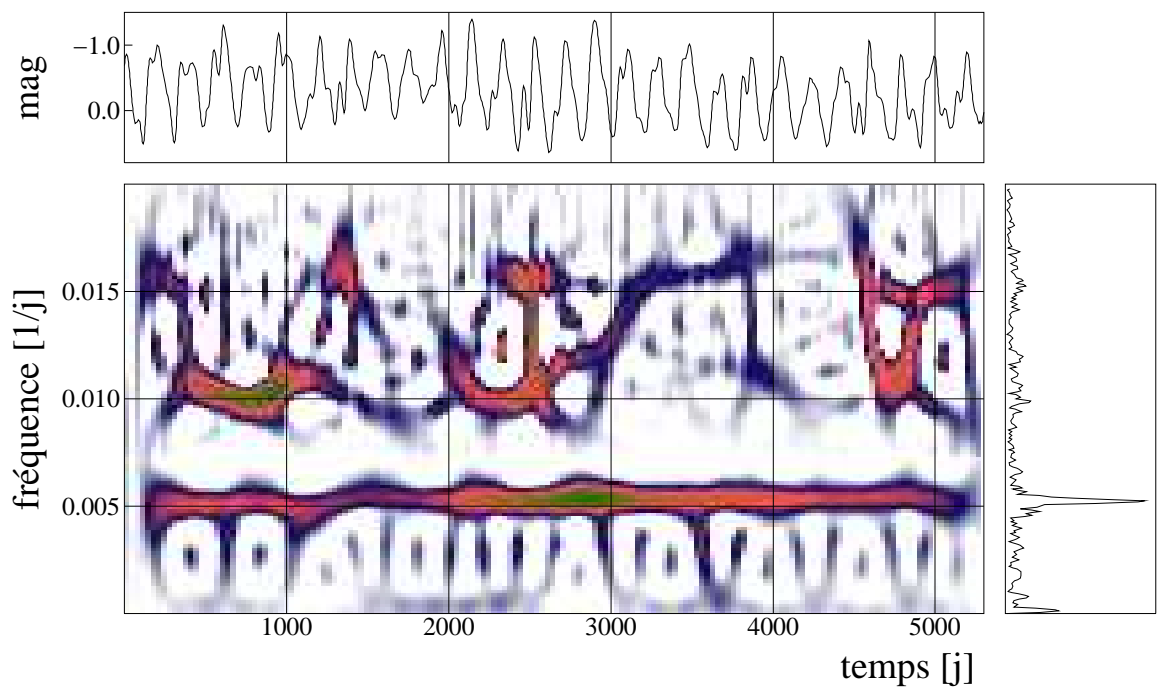}
%    \epsfbox{vcvn2.ps}
 \vskip 1pt
    \caption{En haut: Courbe de lumi\`ere observationnelle liss\'ee de V~CVn;
     \`a droite: spectre de Fourier (amplitude); 
     centre:  diagramme temps-fr\'equence.   }
    \label{vcvn}
  \end{center}
\end{figure}

\begin{figure}[htbp]
  \begin{center}
    \epsfysize=8cm
    \leavevmode
    \epsfbox{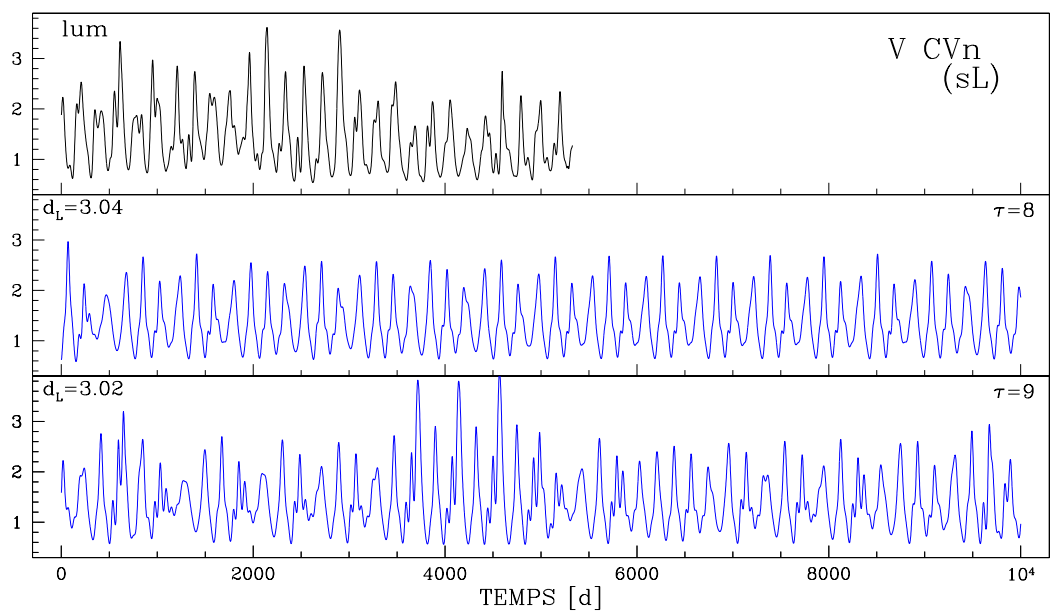}
    \caption{De haut en bas: Courbes de lumi\`ere observationnelle (luminosit\'e)
             liss\'ee de V~CVn; courbe de lumi\`ere synth\'etique typique.  A
             gauche est indiqu\'ee la dimension fractale $d_L$ et \`a
             droite le d\'elai $\tau$ utilis\'e.}
    \label{vcvn_syn}
  \end{center}
\end{figure}

Il semble donc que la dimension physique qui, nous rappelons, est
born\'ee entre $d_L \sim 3.25$ et $d_e^{min}=4$, soit aussi \'egale \`a 4
pour cette \'etoile.

Comme pour AC~Her la courbe de lumi\`ere n'a pas de phases de faible
amplitude.  Par cons\'equent le voisinage du point fixe est mal explor\'e
et il n'est pas possible de confirmer l'existence d'une r\'esonance.

\subsection{L'\'etoile V Canum Venaticorum}

La courbe de lumi\`ere liss\'ee, le spectre de Fourier de l'amplitude
et le diagramme temps-fr\'equence sont montr\'es dans la figure
\ref{vcvn}.

Une reconstruction n'a pas \'et\'e possible avec la courbe de lumi\`ere (en
magnitude) pour cette \'etoile, m\^eme en jouant avec le param\`etre de lissage
$\sigma$, ceci malgr\'e que la courbe de lumi\`ere s'\'etende sur 27 cycles. On
ne peut que sp\'eculer sur la raison de ce manque de succ\`es: peut-\^etre que
la courbe de lumi\`ere n'explore pas assez de son espace de phase, peut-\^etre
que la dynamique est de dimension bien plus grande, peut-\^etre que l'\'etoile
est dans une phase de perte de masse qui l'obscurcit et donne donc lieu \`a des
variations de luminosit\'e superpos\'ees \`a la dynamique propre, ou
peut-\^etre un m\'elange de toutes ces raisons.  Nous pensons cependant que
c'est surtout pour la premi\`ere raison specifi\'ee.  En effet, une
reconstruction \`a 4D, avec les points observationnels convertis d'abord en
magnitude, puis liss\'es, s'av\`ere assez satisfaisante, quoiqu'avec seulement
deux valeurs de $\tau=8$ et 9, et donnant des $d_L=3.01 - 3.5$.  Deux courbes
synth\'etiques typiques sont montr\'ees dans la figure \ref{vcvn_syn}.

\begin{figure}[htbp]
  \begin{center}
    \epsfysize=9cm
    \leavevmode
    \epsfbox{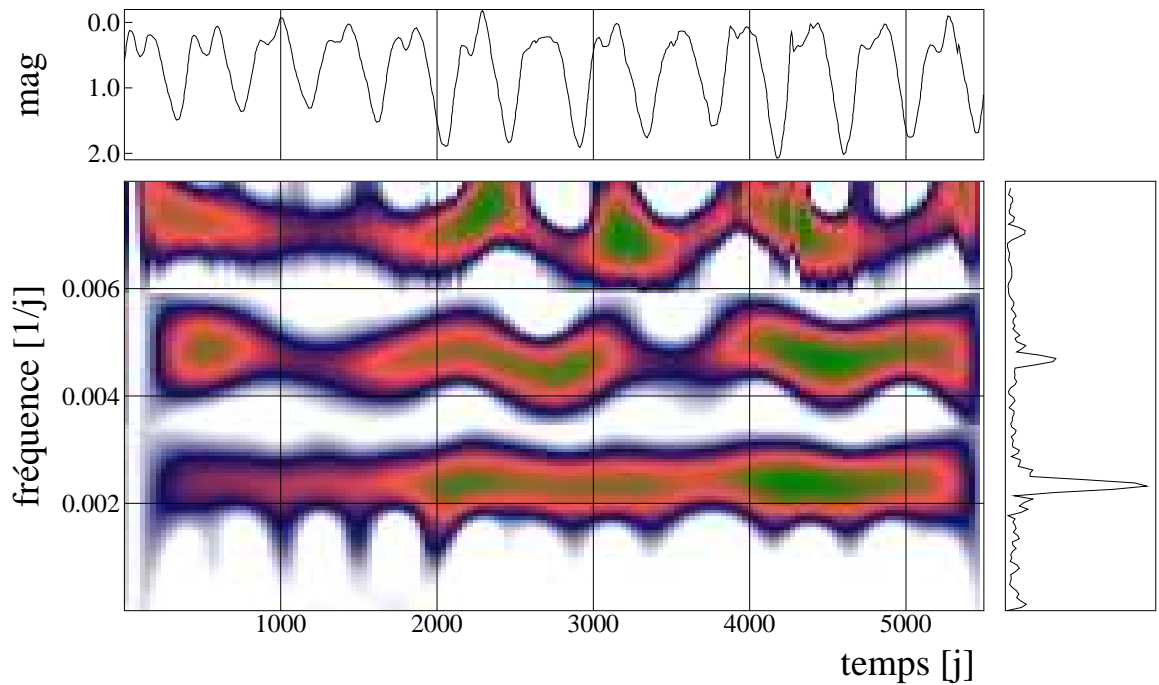}
%    \epsfbox{rscyg.ps}
 \vskip 1pt
    \caption{En haut: Courbe de lumi\`ere observationnelle liss\'ee de RS~Cyg;
     \`a droite: spectre de Fourier (amplitude); 
     centre:  diagramme temps-fr\'equence.   }
    \label{rscyg}
  \end{center}
\end{figure}

\begin{figure}[htbp]
  \begin{center}
    \epsfysize=8cm
    \leavevmode
    \epsfbox{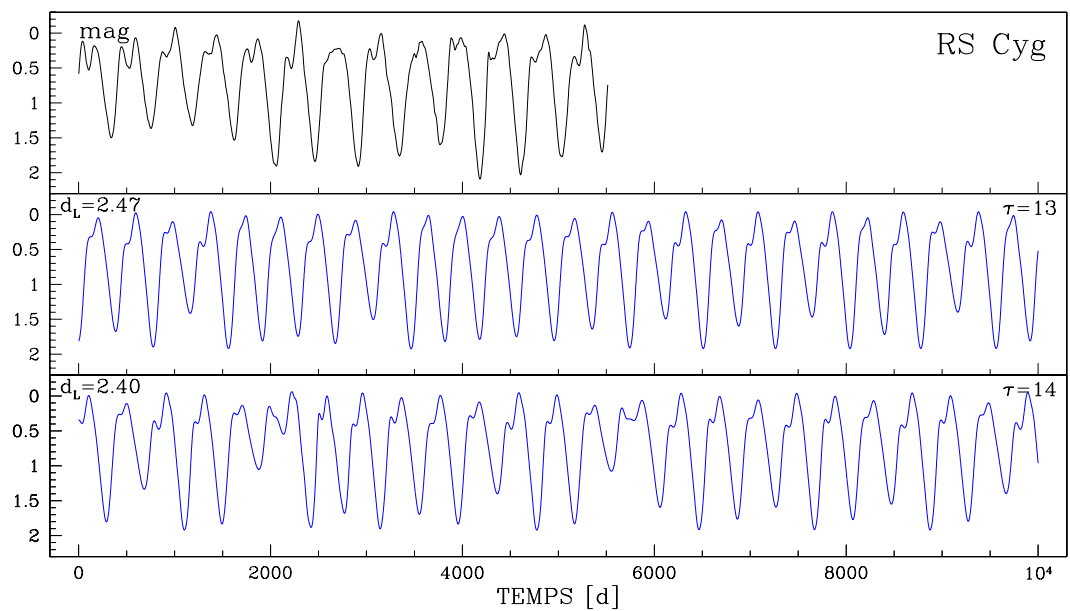}
    \caption{De haut en bas: Courbes de lumi\`ere observationnelle (magnitudes)
             liss\'ee de RS~Cyg; courbe de lumi\`ere synth\'etique typique.  A
             gauche est indiqu\'ee la dimension fractale $d_L$ et \`a
             droite le d\'elai $\tau$ utilis\'e.}
    \label{rscyg_syn}
  \end{center}
\end{figure}

 Nous concluons que les r\'esultats sont tr\`es fragiles pour cette \'etoile,
quoiqu'ils soient consistants avec une dimension 4.

\subsection{L'\'etoile RS Cygni}

Finalement, la courbe de lumi\`ere liss\'ee, le spectre de Fourier de
l'amplitude et le diagramme temps-fr\'equence de RS~Cyg sont montr\'es dans la
figure \ref{rscyg} et quelques courbes synth\'etiques typiques sont montr\'ees
dans la figure \ref{rscyg_syn}.  On note que les observations de la courbe de
lumi\`ere ne s'\'etendent que sur 13 cycles et il ne faut pas s'attendre \`a
des miracles pour la reconstruction.

La reconstruction du flot est relativement robuste avec des d\'elais allant de
$\tau =$ 11 a 17 jours. {\it Prima facie} la comparaison avec les donn\'ees est
tr\`es favorable, mais une inspection plus en profondeur montre que la
reconstruction a un peu du mal \`a capturer d\^ument les alternances (en forme
de dent molaire) tr\`es typiques des \'etoiles RV~Tau.  Les dimensions $d_L$
varient de 2.1 \`a 3.1.  Nous trouvons qu'une condition de r\'esonance 2:1 a
l'air d'\^etre satisfaite pour $\tau=$ 11, 12 et 15.

Les reconstructions \`a 3D ne sont pas satisfaisantes du tout ce qui implique
que la dimension de plongement minimale est plus grande que 3.  A 5D
la densit\'e de points est trop faible pour une reconstruction robuste, mais
les r\'esulats \`a 4D montr\'es dans la figure \ref{rscyg_syn} sugg\`erent
$d_e^{min} = 4$.  Nous pensons que la dimension physique $d=4$ pour pouvoir
accommoder deux modes vibratoires (\`a proprement parler on ne peut pas
\'eliminer la possibilit\'e $d=4$ si $d_L<3$ \`a cause de $d_L < d \le
d_e^{min}$).

On en conclut que l'analyse de RS~Cyg est prometteuse, mais qu'une
reconstruction fiable requiert une base de donn\'ees observationnelles plus
importante.

\section{Conclusion}

La m\'ethode de reconstruction de flot a \'et\'e appliqu\'ee aux courbes de
lumi\`ere irr\'eguli\`eres observationnelles de cinq \'etoiles du type
RV~Tau et une \'etoile semi-r\'eguli\`ere.  Les p\'eriodes de ces \'etoiles
varient de $\sim$ 35 jours dans le cas de AC~Her \`a $\sim$ 400
jours pour RS~Cyg.
 
La qualit\'e de la reconstruction de flot est variable.  
et un sommaire des r\'esultats est donn\'e dans le tableau 1.

\vskip 9mm

\begin{tabular}{||c||c|c|c|c|c|c|c||}
\hline
                                  & & & & & & \\
\color{blue}\quad \quad Etoile \quad \quad           &\color{blue} \quad\quad $\tau$
\quad\quad &\color{blue} \quad \quad $d_L$ (\`a 4D)\quad \quad   
    &\color{blue}\quad \quad  $d_e^{min}$ \quad \quad 
& \color{blue}\quad $\omega_2/\omega_1$ \quad &\color{blue}
 \quad \`a  5D \quad  & \color{blue}\quad \quad  $d$ \quad \quad  \color{black} \\
                                  &              &          & & &
				  &     \\
\hline
                                  &              &          & & &
				  &     \\
\color{red}R Sct\color{black}    & 4 -- 8    & 3.05 -- 3.2 &4 & 2.13          
& $d_L < 4$ & 4 \\
\color{red}AC Her\color{black}    & 3 --13   & 2.05 -- 2.9 &3, 4, 5 & --            
& $d_L < 3$ & 3, 4, 5\\
\color{red}SX Her\color{black}    & 7, 9     & 3.1 -- 3.8  &4 & $\sim$2.15   
& $d_L < 4$ & 4\\
\color{red}SX Her-L\color{black}  & 7, 8     & 3.2 -- 3.4  &4  & $\approxgt 2$ 
& --        & 4 \\
\color{red}R UMi\color{black}     & 8 --14   & 3.1 -- 3.4   &4 & --            
& $d_L < 4$ & 4\\
                                  & 3 --6    & 2.05 -- 2.4  &4 & --            
				  & -- & 3, 4 \\
\color{red}V CVn\color{black}     & 8, 9     & 3.01 -- 3.5  &4 & --            
& --        & 4\\
\color{red}RS Cyg\color{black}    & 11 -- 17 & 2.1 -- 3.1   &4 &$\approxgt 2$  
& --        & 3, 4\\
\hline
\end{tabular}

\vskip 8mm

\centerline{Table:  Sommaire.}

\vskip 3mm

La seconde colonne indique les valeurs du d\'elai pour lequelles une
reconstruction robuste et stable existe (stable dans le sens que des signaux
synth\'etiques de plus de 10000 jours existent et ressemblent au signal
observationnel).  La troisi\`eme colonne donne les valeurs des dimensions
fractales obtenues avec les signaux synth\'etiques (sauf dans le cas de R~Sct
la variation de $d_L$ est large \`a cause de la faible longueur des signaux
synth\'etiques).  La 4e colonne donne la dimension de plongement minimale
indiqu\'ee par nos reconstructions, et la 5e colonne les rapports de
fr\'equence des racines spirales du point fixe.  La colonne 6 indique les
valeurs de la dimension fractale quand une reconstruction \`a 5 D a \'et\'e
possible.  Finalement la derni\`ere colonne r\'esume la dimension physique de
la dynamique stellaire qu'on tire des bornes (eq. 6).
Pour R~Sct la reconstruction est tr\`es robuste, elle est presqu'aussi robuste
pour SX~Her, alors qu'elle l'est tr\`es peu pour V~CVn et AC~Her, et elle est
interm\'ediaire pour les deux autres \'etoiles.  Ce qui est important cependant
c'est que nos r\'esultats, moins convainquants quand pris s\'epar\'ement, sont
en accord avec, et donc corroborent ceux qu'on a trouv\'es avec R~Sct.

Ceci sugg\`ere que le m\'ecanisme sous-jacent de toutes ces \'etoiles est
probablement le m\^eme:  l'espace de phase physique est de
dimension quatre.  Dans cet espace un mode vibratoire qui est (lin\'eairement)
tr\`es instable, et a donc tendance \`a cro\ih tre, entra\ih ne par r\'esonance
2:1 un mode (lin\'eairement) stable qui lui a tendance \`a d\'ecro\ih tre.
Dans un sc\'enario \`a la Shilnikoff (g\'en\'eralis\'e) ce couplage
nonlin\'eaire donne lieu \`a une oscillation chaotique.

Le fait que les pulsations de ce type d'\'etoile soient chaotiques et de faible
dimension peut surprendre \`a cause de leur nature physique assez violente,
donnant lieu \`a des variations de luminosit\'e allant jusqu'\`a des facteurs
40 dans le cas de R~Sct.  Cependant on note que des calculs hydrodynamiques
avaient pr\'edit l'existence de chaos dans ce type d'etoile \cite{b87}
\cite{pdchaos88} et une analyse topologique de ces r\'esultats th\'eoriques
l'avait corrobor\'e \cite{letellier}.

La faible dimension peut aussi surprendre \`a cause de la pr\'esence certaine
de turbulence et de convection dans ces \'etoiles, alors qu'il est bien connu
que ces ph\'enom\`enes physiques sont de haute dimension.  Que nous ayons
r\'eussi \`a extraire la dynamique de ces donn\'ees observationnelles est sans
doute d\^u au fait que les \'echelles de la pulsation et de de la turbulence
sont diff\'erentes.  De plus les amplitudes des mouvements turbulents sont
tr\`es faibles par rapport \`a celle de la pulsation propre et ont \'et\'e
largement supprim\'ees dans le lissage des donn\'ees.

En perspective il est int\'eressant de noter que la m\'ethode de reconstruction
de flot, combin\'ee aux calculs hydrodynamiques a donc permis de r\'esoudre le
vieux myst\`ere concernant la nature irr\'eguli\`ere des pulsations de ces
\'etoiles.

\vskip 20pt

L'un de nous (JRB) remercie Christophe Letellier de nous avoir fourni les
moyens de participer \`a cet atelier.  Notre recherche a profit\'e du support
financier de la NSF ainsi que de l'Acad\'emie Hongroise des Sciences.


\vfill

\begin{thebibliography}{let1}


\bibitem{cox}
Cox, J. P. {\it Theory of Stellar Pulsation}, Princeton Series in Astrophysics
(1980).

\bibitem{buchler93}
 Buchler, J. R. in {\sl Nonlinear Phenomena in Stellar Variability},
     Eds. M. Takeuti \& J.R. Buchler, {Dordrecht: Kluwer Publishers}, reprinted
     from 1993, Ap\&SS, 210, 1--31 (1993).\hfill\break
     [http://www.phys.ufl.edu/$\sim$buchler/mito.ps.gz]

\bibitem{arp55}
Arp, H. C., {\it Astron. J.} 60, 1--11 (1955).

\bibitem{b87}
Buchler, J.R. \& Kov\'acs, G., 
   %{\sl Period-Doubling Bifurcations and
   %  Chaos in W Virginis Models}, 
   {\it Astrophys. J. Lett}. 320,
     L57--62 (1987).

\bibitem{pdchaos88}
 Kov\'acs G. \& Buchler J.R., 
 %{\sl Regular and Irregular Pulsations in   Population II Cepheids}, 
{\it Astrophys. J.} 334, 971--994
(1988).

\bibitem{aik90}
Aikawa, T.,  {\it Astrophys. \& Space Sci.} 164, 295--308 (1990).

\bibitem{bgk87}
Buchler, J. R., Goupil, M. J. \& Kov\'acs, G., {\it Phys. Lett}, A126, 
177--180 (1987).

\bibitem{aik87}
Aikawa, T., {\it Astrophys. \& Space Sci.}, 139, 281--293 (1987).

\bibitem{pol96}
Pollard, K. et al., "MACHO Observations of Type II Cepheids and RV Tauri Stars
in the LMC", in {\it Variable Stars and the Astrophysical
     Returns of Microlensing Surveys}, Eds. R. Ferlet, J.P. Maillard \&
     B. Raban, Editions Fronti\`eres, pp. 219--223, (1996).

\bibitem{rsctprl}
Buchler, J. R., Serre, T., Koll\'ath, Z. \& Mattei, J. 
%{\sl A     Chaotic Pulsating Star -- The Case of R~Scuti}, 
{\it Phys. Rev. Lett.}
      74, 842--845 (1995).

\bibitem{rsct96}
Buchler, J. R., Koll\'ath, Z., Serre, T. \& Mattei, J.,
 % {\sl A Nonlinear Analysis of the Variable Star R Scuti}, 
{\it Astrophys. J.}
      462, 489--504
(1996).

\bibitem{acher98}
Koll\'ath, Z., Buchler, J. R., Serre, T. \& Mattei, J.,
% {\sl      Analysis of the Irregular Pulsations of AC~Herculis}, 
 {\it Astron. \& Astrophys.} 329, 147--155
(1998).

\bibitem{potsdam}
Buchler, J. R., Koll\`ath, Z. \& Cadmus, R., Proceedings of Chaos 2001,
Potsdam, Germany (2001) (in press).

\bibitem{reinsch}
Reinsch, C. H., {\it Numerische Mathematik}, 10, 177--201 (1967).

\bibitem{letellier-floride} 
Letellier, C., Le Sceller, L., Gouesbet, G., Brown, R., Buchler, J.R. \&
      Koll\'ath, Z., {\sl Nonlinear Signal and Image Processing}, 
      Eds. J. R. Buchler \& H. Kandrup, Ann.
      N.Y. Acad. Sciences, 808, 25--50 (1996).

\bibitem{serre}
Serre, T. \& Nesme-Ribes, E.,
Astronomy and Astrophysics, 360, 319--330 (2000)

\bibitem{weigend}
  Weigend, A.S \& Gershenfeld, N. A.,
  {\it Time Series Prediction} (Addison-Wesley: Reading) (1994).

\bibitem{serre96}
Serre, T., Koll\'ath, Z. \& Buchler, J. R., 
% 1996, 
%{\sl Search For
%     Low--Dimensional Nonlinear Behavior in Irregular Variable Stars -- The
%     Global Flow Reconstruction Method}, 
{\it Astron. \& Astrophys.},
     311, 833--844 
(1996).

\bibitem{varenna}
Buchler, J.R., "Search for Low-Dimensional Chaos in
     Observational Data", International School of Physics "Enrico Fermi",
     Course CXXXIII, {\it Past and Present Variability of the
     Solar-Terrestrial System: Measurement, Data Analysis and Theoretical
     Models}, Eds. G. Cini Castagnoli \& A. Provenzale, pp. 275--288, 
     Societ\`a Italiana de Fisica, Bologna, Italy (1996).\hfill\break
     [http://xxx.lanl.gov/abs/chao-dyn/9707012]

\bibitem{takeuti}
Buchler, J. R. \& Koll\'ath, Z., "Nonlinear Analysis of
     Irregular Variables", in {\it Nonlinear Studies of Stellar Pulsation},
     Eds. M. Takeuti \& D.D. Sasselov, {\it Astrophy. \& Space Sci. Libr.
     Ser.} 257, 185--213 (2001)
     [http://xxx.lanl.gov/abs/astro-ph/0003341].

\bibitem{cohen}
Cohen, L. 1994, Time-Frequency Analysis. Prentice-Hall PTR. Englewood Cliffs,
NJ

\bibitem{kollath90}
Koll\'ath Z., {\it Month. Not. Roy. Astron. Soc.}, 247, 377--386 (1990).

\bibitem{ott}
Ott, E., {\it Chaos in Dynamical Systems}, Cambridge Univ. Press
(1993).

\bibitem{tresser}
Glendenning, P. \& Tresser, C., {\it J. Physique Lett.} 46, L347--352 (1985).


\bibitem{wvir} 
Serre, T., Koll\'ath, Z. \& Buchler, J. R.,
 % {\sl Search For
 %    Low--Dimensional Nonlinear Behavior in Irregular Variable Stars -- The
 %    Analysis of W~Vir Model Pulsations}, 
 {\it Astron. \& Astrophys.},
     311, 845--851 (1996).


\bibitem{brown}
Kennel, M.B., Brown, R. \& Abarbanel, H.D.I., Phys. Rev. A45, 3403 (1992)


\bibitem{letellier} 
 Letellier, C., Gouesbet, G., Soufi, F., Buchler, J.R.\& Koll\'ath,
     Z. 
% {\sl Chaos In Variable Stars~: Topological Analysis of W Vir
%     Model Pulsations}, 
{\it Chaos}, 6 (3), 466--476 (1996).


\end{thebibliography}
\end{document}